\documentclass[aps]{revtex4}
\usepackage[english]{babel}
\usepackage{graphicx}

\begin{document}
\title{Electroweak interactions and the muon $g-2$: bosonic two-loop effects}
\author{ T. Gribouk}\email{tgribouk@phys.ualberta.ca}
\author{A. Czarnecki}
\affiliation{Department of Physics, University of Alberta, Edmonton,
AB, Canada $T6G  2J1$}
\begin{abstract}
   We present a detailed evaluation of
   the bosonic two-loop electroweak corrections to the muon's anomalous magnetic moment.
   We study the Higgs mass dependence and find agreement with a previous
   evaluation in the large Higgs mass limit.
   We find $a_{\mu}^{\rm EW\ bos}({\mbox{two-loop}})=(-22.2 \pm
   1.6)\times 10^{-11}$, for $114 {\rm \ GeV} \le M_{\rm Higgs}\le 700 {\rm \ GeV}$.
\end{abstract}
\maketitle

 \rm
\section{Introduction}
\label{intro} The anomalous magnetic moment of the muon,
$a_{\mu}\equiv\left( g_{\mu}-2\right)/2$ has recently been
determined with a very high precision. A series of measurements with
positive and negative muons by the E821 Collaboration at the
Brookhaven National Laboratory resulted in the present average value
\cite{Bennett:2004pv}
\begin{eqnarray}
\label{g-2Res} a_{\mu}^{\rm exp}=(116\,592\,080 \pm 60)\times
10^{-11}.
\label{exp}
\end{eqnarray}
At this level of a one half part per million precision, this quantity is
sensitive to subtle effects predicted by the Standard Model (SM), including
five-loop quantum electrodynamics (QED), hadronic vacuum polarization and
light-by-light scattering, and electroweak interactions at two-loops.

Obviously, $g_\mu-2$ may be affected also by interactions beyond
the SM.  For this reason, many researchers have analysed $g_\mu-2$
in various models of new physics and performed sophisticated
studies of the SM effects which are an irreducible background in
the search of unknown phenomena.  A summary of the SM prediction
can be found, for example, in the recent studies and reviews
\cite{Davier:2004gb,Nara,Melnikov:2003xd,Passera:2004bj}.

At present, the experimental result in Eq. (\ref{exp}) exceeds the
SM prediction by about 2.6 times the combined theoretical and
experimental uncertainty.  This tantalizing discrepancy may be due
to an effect of a new interaction, perhaps supersymmetry. However,
it is important to scrutinize the SM prediction before a conclusion
can be made. The present paper is devoted to a reevaluation of the
largest part of two-loop electroweak diagrams, namely those without
closed fermionic loops.

Electroweak one-loop corrections to $g_\mu-2$ were among the first
quantum effects studied in the renormalizable electroweak theory, in
1972  \cite{fls72,Jackiw72,ACM72,Bars72,Bardeen72}.  They were found
to increase $g_\mu-2$ by
\begin{eqnarray}
 a_\mu ^{EW}
 \left(
    {\rm 1 loop}
 \right)
 = \frac{5}{3}\frac{{G_\mu m^2 }}{{8\sqrt 2 \pi ^2 }} \times
 \left[
    {1 + \frac{{
    \left(
        {1 -4\sin ^2 \theta _w }
    \right)
    ^2 }}{5} + O
    \left(
        {\frac{{m^2}}{{M^2 }}}
    \right)
    }
 \right]
 \approx 195 \times 10^{ - 11}\,,
\label{oneLoop}
\end{eqnarray}
where $G_{\mu}=1.16639(1)\times 10^{-5} GeV^{-2}$  and weak mixing
angle $\sin^2\theta _w=1-M_W^2/M_Z^2$.  The large mass parameter $M$
represents the mass of a $W$, $Z$, or a Higgs boson.

This effect was too small to be measured in the then ongoing CERN
experiment.  The desire to observe it motivated in part the latest
Brookhaven effort.

Two decades after the first electroweak result, Kukhto et al.
\cite{KKSS} found that an additional virtual photon may
significantly modify the one-loop value in Eq. (\ref{oneLoop}). They
estimated that effect as a -22\% reduction -- surprisingly large for
hard virtual photons.  This and an analogous reduction of the rare
muon decay $\mu\to e\gamma$ \cite{Czarnecki:2001vf} are due to the
large anomalous dimension of the dipole operators such as $\bar \mu
\sigma^{\mu \nu}\mu F_{\mu\nu}$ and $\bar e \sigma^{\mu \nu}\mu
F_{\mu\nu}$.

That large effect found in a subset of two-loop electroweak
contributions was similar in size to the design precision of the
E821 experiment.  It therefore appeared as warranted, even
necessary, to evaluate the complete two-loop result  -- a
calculation that had not been performed before in the electroweak
theory for any other observable.

In 1995 a complete set of 1678 electoweak two-loop diagrams for
$g_\mu-2$ was generated by a computer system \cite{Kaneko:1995kr}.
However, such large number of diagrams, many of which are divergent,
could not be numerically calculated, at least at that time.
Fortunately, it turned out that the majority of those diagrams are
strongly suppressed by extra factors of the
muon-to-intermediate-boson mass ratio and can be neglected.  Thus, a
complete two-loop result was found \cite{CKM96}.  The numerical
value was found to be dominated by large logarithms arising due to
photon exchanges and was, somewhat accidentally, close to the value
found in its first studies \cite{KKSS,KKS}.

In later work, the renormalization group equation was employed to
estimate higher-order logarithmic effects, which however are not
sizable \cite{Czarnecki:2002nt,Degrassi:1998es}.

The result of \cite{CKM96} was obtained in an approximation assuming
that the Higgs boson is sufficiently heavier than the $W$ and $Z$
bosons such that $M_{W,Z}/M_H$ can be used as an expansion
parameter.  A recent study \cite{Heinemeyer:2004yq} relaxed that
approximation and found the value of two-loop contributions valid
also for a light Higgs (moreover, the two-loop supersymmetric
effects were evaluated in that paper).

In this paper, we reanalyse the two-loop electroweak effects.  Our
goal is to check the previous results and study the Higgs mass
dependence. We present our results in a semi-analytical form. That
is, the dependence on the Higgs mass is presented analytically,
while some parts of expressions that depend only on the well-known
particle masses are, for the sake of brevity, evaluated numerically.
A method employed for obtaining analytical results is also described
in some detail.

In this work we focus on diagrams with only bosonic loops (no closed
fermion loops).  The fermionic subset of corrections was studied
separately \cite{CKM95,Peris:1995bb,Knecht:2002hr}.  It was recently
subject of an interesting theoretical controversy which seems to be
settled now (see \cite{Czarnecki:2002nt} for a thorough discussion
and references).

In the next Section we briefly explain the asymptotic operation
which is the main technical tool used to obtain the analytical
result for $g_\mu-2$.  Section \ref{SMcontribution} presents partial
results for various groups of contributing diagrams. First, we
divide all diagrams into five subsets of topologically equivalent
diagrams. Then we give analytical or semi-analytical result for each
topology. Finally we discuss the renormalization procedure and
evaluate necessary counterterms, provide the final numerical result
and compare it with the results of Ref.
\cite{CKM96,Heinemeyer:2004yq}.

\section{Techniques: Asymptotic operation and resulting integrals}
\label{techniques}

The method of asymptotic operation (see \cite{Tkachev:1994gz} for a
review and references) has been an invaluable tool in numerous
recent studies of effects involving various energy scales. In the
present problem, the mass of the muon $m$ sets the soft scale, and
the masses of the $W$, $Z$, and Higgs bosons -- the hard scale:
$M_{W,Z,H}\gg m$.   As we explained in the Introduction, the
original study \cite{CKM96} assumed in addition $M_H \gg M_{W,Z}$.
Here, we will not assume any hierarchy between $M_{W,Z}$ and $M_H$
at the price of certain complication of our results.

We would like to explain here the basic principles of asymptotic
operation.  Instead of attempting a rigorous derivation or even a
rigorous exposition, we will use a simple example to elucidate the
method.

In this study, we are interested in two-loop Feynman graphs $G$,
which have a soft-scale ($\sim m$) external momentum and involve
internal lines with both soft-scale and  hard-scale ($\sim M$)
virtual particles. The exact value of the two loop integrations is a
(possibly very complicated) function of $m/M$. However, for our
purposes it is entirely sufficient to know only an expansion of that
function up to $m^2/M^2$.  The purpose of the asymptotic operation
is to obtain the desired order of that expansion without having to
compute the whole function.

For our purposes, the action of the asymptotic operation on a
Feynman graph $G$ may be described with the following formula,
\begin{eqnarray}
\label{GENas}
 {\rm As} \circ G =
% \tau  \circ G +
 \sum\nolimits_{
% \left\{
    {h }
% \right\}
 }
 \left(
    {%\prod\nolimits_i
    {\tau  \circ h } }
 \right)
 \times
 \left[
    {G\backslash %\prod\nolimits_i
    {h } }
 \right].
\end{eqnarray}
Let us first of all explain the notation.  $ {\rm As} \circ G$ is an
expansion of the exact value of $G$ in powers {\em and logarithms}
of $m/M$.  $h$ are the subgraphs of $G$ in which all loop momenta
are considered as hard.

\begin{figure}[hbt]
\begin{tabular}{c@{\hspace*{7mm}}c}
    \begin{tabular}{c}
    \\
        \scalebox{0.9}
        {\includegraphics[width=.35\textwidth]{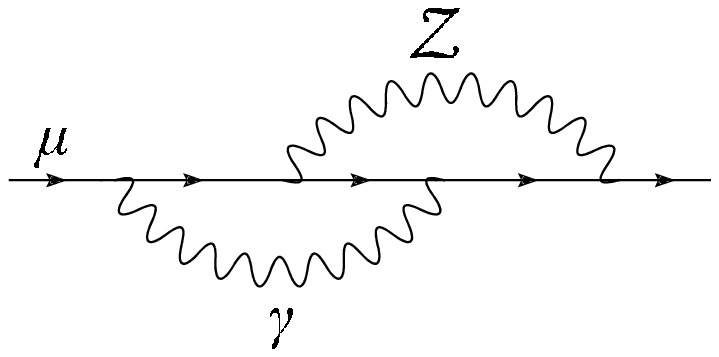}}
    \\
        (a)
    \end{tabular}
&
    \begin{tabular}{c@{\hspace*{7mm}}c}
        \scalebox{0.6}
        {\includegraphics[width=.35\textwidth]{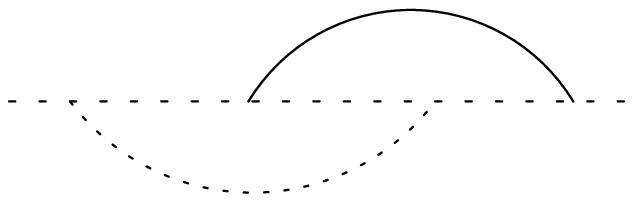}}
    &
        \scalebox{0.6}
        {\includegraphics[width=.35\textwidth]{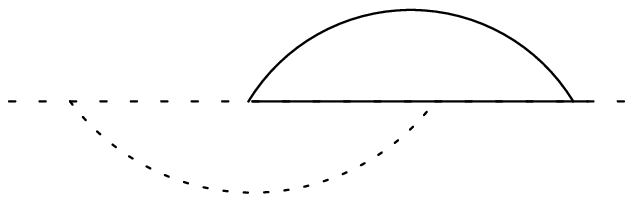}}
    \\
        (b)
    &
        (c)
    \\
    \\
        \scalebox{0.6}
        {\includegraphics[width=.35\textwidth]{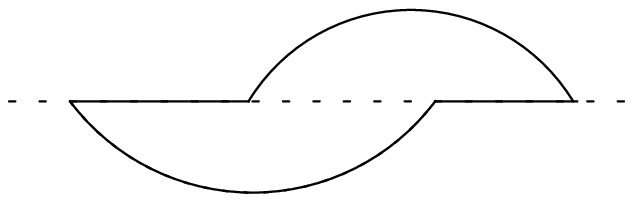}}
    &
        \scalebox{0.6}
        {\includegraphics[width=.35\textwidth]{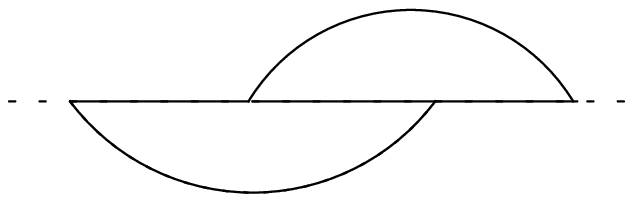}}
    \\
        (d)
    &
        (e)
    \end{tabular}
\\
\end{tabular}  \\
\caption{An example of applying the asymptotic operation to a graph $G$,
shown in (a).  (b)-(e) show the full set of subgraphs, as described in the text.
Propagators through which hard momenta flow are indicated with solid lines -- they constitute
a subgraph $h$.  Dashed lines contain only soft-scale momenta and masses -- they constitute $G\backslash h$.}
 \label{asympt}
\end{figure}

Instead of giving an exact definition of which $h$ are relevant,
we use as an example the diagram shown in Fig. \ref{asympt}.
 In
that Figure, $G$ is shown in picture (a). Subgraphs $h$ are shown
in (b)--(e), and range from a single heavy particle line to the
whole graph.   $\tau$ is the Taylor expansion operator, expanding a
subgraph in all soft-scale parameters such as the light-particle
masses and external momenta of that subgraph (the latter include
loop momenta of those loops that are not part of the subgraph).
Finally, $ {G\backslash {h } }$ denotes the graph $G$ with the
subgraph $h$ contracted into a point.  The physical intuition for
that concept is such that all masses and loop momenta in $
{G\backslash {h } }$ are soft; thus, from their point of view, the
hard-scale subprocess  in $h$ occurs at a very short-distance
scale and can be contracted to a point. In other words, the
subgraphs $h$ create effective vertices.

What do we gain from this construction?  We completely separate soft
and hard scales for the purposes of the integration.  Thus, the
resulting integrals cannot contain both $m$ and $M$ -- they
factorize into a part that depends only on the soft scales and
another that depends on the hard ones.  No non-trivial functional
dependence on $m/M$ can arise and we obtain the expansion to a given
order in $m/M$ by taking a sufficient number of terms in the Taylor
expansion of the subgraphs.

\begin{figure}[hbt]
\begin{tabular}{c@{\hspace*{20mm}}c@{\hspace*{20mm}}c}
        {\includegraphics[width=.35\textwidth]{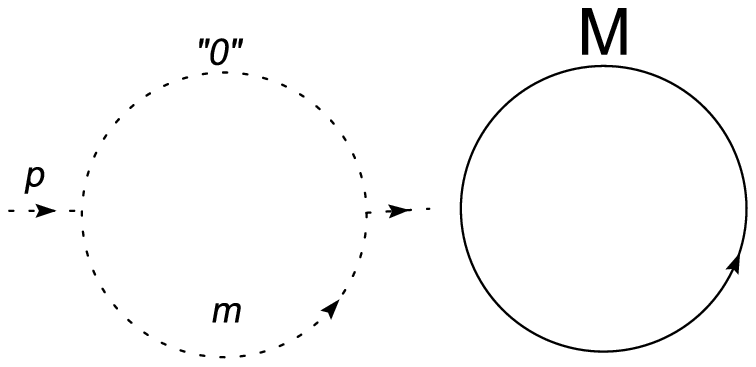}}
&        \scalebox{0.5}
        {\includegraphics[width=.35\textwidth]{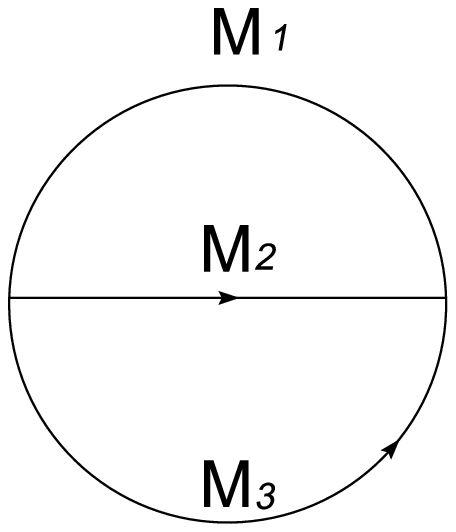}}
 &       \scalebox{0.5}
        {\includegraphics[width=.45\textwidth]{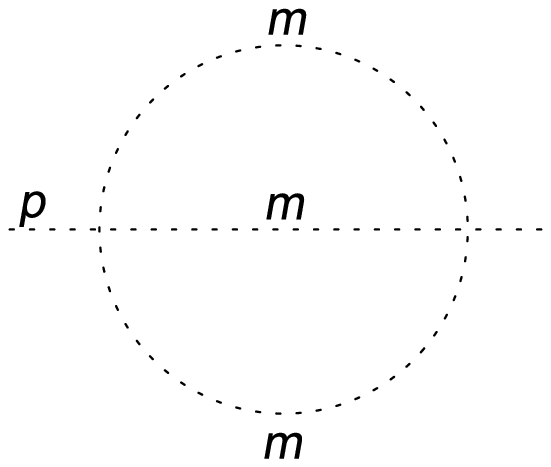}}
        \\
        (a) & (b)& (c)
  \end{tabular}  \\
\caption{Basic types of loop integrals needed for the evaluation of
two-loop electroweak corrections to $g_\mu-2$: (a) One-loop
light-mass on-shell integral multiplied by a one-loop  vacuum
integral with a large mass; (b) Two-loop vacuum integral (some
propagators may be massless $(M=0)$); (c) Two-loop light-mass
on-shell integral.} \label{types}
\end{figure}

The asymptotic operation greatly simplifies the types of integrals
that we need to evaluate, as shown in Fig. (\ref{types}).
 One-loop on-shell integrals and one-loop massive integrals,
pictured in Fig. (\ref{types})(a), are trivial.

Two-loop integrals in Fig. (\ref{types})(b), dependent on arbitrary
masses, were computed in Ref. \cite{adt93,inlo15}. By means of
integration by parts identities  \cite{Tkachov:1981wb} the initial
integral with arbitrary powers of propagators can be reduced to the
integral with all propagators in power one. For the case when all
three propagators are massive, we encounter only diagrams with two
different masses present. In this simplified case, the general
result for such integral, in the dimensional regularization, becomes
\begin{eqnarray}
\label{M3int}
 \ {{I_{m m M} }} &\equiv&
 \frac{\left(4\pi\right)^d}{\Gamma^2(1+\varepsilon)}\int\int\frac{{\rm d}^d p ~{\rm d}^d
 q}{\left(p^2+m^2\right)\left(q^2+m^2\right)\left(\left(p+q\right)^2+M^2\right)}
 \nonumber\\&= &
 \frac{\left(m^2\right)^{1-2\varepsilon}}{\left(1-\varepsilon\right)\left(1-2\varepsilon\right)}
 \left[
    -\frac{1+2z}{\varepsilon^2}+\frac{4z \ln 4z}{\varepsilon}-2z \ln^2 4z+2(1-z)\Phi(z)
 \right]\,,
\end{eqnarray} where $d=4-2\varepsilon$,
 $z=\frac{M^2}{4m^2}$, and  the function
$\Phi$ is defined as
\begin{eqnarray}
\label{Phi1} \Phi(z)= 4
 \left(
    \frac{z}{1-z}
 \right)
 ^{\frac{1}{2}} {\rm Cl_2}
 \left[
    2\arcsin
    \left(
        z^{\frac{1}{2}}
    \right)
 \right]
 ,\hspace{10mm}
 {\rm Cl_2}(\theta)=-\int_{0}^{\theta}d\theta \ln|2\sin(\theta/2)|
 \,,
 \end{eqnarray}
for $z<1$, and
\begin{eqnarray}
\label{Phi2} \Phi(z)=
 \left(
    \frac{z}{z-1}
 \right)
 ^{\frac{1}{2}}
 \left\{
    -4{\rm Li_2}(\xi)+2 {\rm ln}^2\xi-{\rm ln}^2 4z +\frac{\pi^2}{3}
 \right\}
 ,\hspace{10mm}
 \xi= \frac{1-
 \left(
    \frac{z-1}{z}
 \right)
 ^\frac{1}{2}}{2}
    \,,
\end{eqnarray} for $z>1$.

 For the case when one propagator is massless we introduce the function
 $I^{\rm fin}$, which is  the finite
 part of the integral $I_{M_c M_a M_b } $ for the case of
 $M_c=0$; denoting $R=\frac{M_b^2}{M_a^2}$ we have
\begin{eqnarray}
\label{M2int} \ I^{\rm fin}(R)=
-{7\over 2} (1+R)+3R\ln R
-(1-R){\rm Li}_2(1-R) -{R\over 2}\ln^2 R.
\end{eqnarray}
Finally, two-loop on-shell integrals shown in Fig. (\ref{types})(c)
can be found, for example, in Ref. \cite{bielefeld}.

\begin{figure}[!h]
    \centering
    \includegraphics[width=0.2\textwidth]{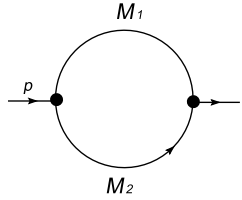}
    \caption{One-loop massive on-shell self energy integral}
    \label{SE}
\end{figure}

To calculate the renormalization counter terms we shall also need
one-loop massive on-shell self-energy integrals shown in Fig.
(\ref{SE}).
They are expressed as linear combinations of integrals of the type
(we use $p^2=-M^2$):
\begin{eqnarray}
\label{SEint} \ S^{a b} (M_1, M_2, M)= \frac{\Gamma
(b-2+\varepsilon)}{\Gamma(b) \Gamma(1+\varepsilon)} \int_0^1 {\rm d}
x \frac{x^a}{\left[(1-x)(M_1^2-xM^2)+x
M_2^2\right]^{(b-2+\varepsilon)}}\,.
\end{eqnarray}
These integrals can be represented as series in $\varepsilon$ and
can be easily computed to the necessary order. We  introduce special
functions $S_{\rm fin}^{a b} (M_1, M_2, M)$ which represent the
finite parts of  (\ref{SEint}). We shall need only the following
cases,
\begin{eqnarray}
\label{S01} S_{\rm fin}^{0 1}(M_1, M_2, M)&=& \int_0^1{\rm d}x
\left\{ \left[(1-x)\left(M_1^2-xM^2\right)+xM_2^2\right] \left[\ln
\left((1-x)\left(M_1^2-xM^2\right)+xM_2^2\right)-1\right]\right\}\,,
\nonumber \\
\label{S02} S_{\rm fin}^{a 2}(M_1, M_2, M)&=&-\int_0^1 {\rm d}x
x^{a}\ln\left[(1-x)\left(M_1^2-xM^2\right)+xM_2^2\right]
    \,,
\end{eqnarray}
with $a$ taking one the values $0, 1$ or $2$.

\section{Bosonic electroweak two-loop contibutions in the SM}
\label{SMcontribution} In this section we apply the technique of
asymptotic operation, as described in Sec.~\ref{techniques}, to
evaluate all two-loop bosonic contributions.  We divide the diagrams
into subsets of various topologies and mass assigments, and  present
detailed results for each subset.  We display the dependence on the
Higgs mass in the analytical form. We also provide details of the
renormalization procedure.

Following \cite{CKM96}, we use the 't Hooft-Feynman non-linear gauge
\cite{Fujikawa:1973qs}. We choose it in such way that we
eliminate the vertex $\gamma
W^ \pm G^ \pm $. Since $g_\mu-2$ involves an external photon in
every diagram, such choice greatly reduces the number of diagrams.
The set of diagrams we have to consider contains all two-loop
diagrams one can compose in $SM$ with the exclusion of pure QED
diagrams and diagrams with closed fermion loop; we drop all diagrams
with more than one scalar coupling to the muon line, since each such
coupling introduces an extra factor $\frac{{m_\mu }}{{M_W}}$. Taking
advantage of the mirror symmetry we reduce the number of diagrams
that must be individually calculated to 138; in addition, some
one-loop diagrams must be evaluated for the renormalization.

\subsection{Two-loop topologies and their evaluation}

\begin{figure}[!h]
    \begin{tabular}{c}
            \begin{minipage}[c]{.45\textwidth}
                \centering
                \begin{tabular}{|c|c|c|c|c|c|}
                        \hline
                        ${\rm Topology}$&$M_A$&$M_B$&$M_C$&$M_D$&$M_E$
                        \protect\\
                        \hline
                        $1_A$&$M_Z$&$M_Z$&$m$&$m$&$m$
                        \protect\\
                        \hline
                        $1_B$&$0$&$M_Z$&$m$&$m$&$m$
                        \protect\\
                        \hline
                        $1_C$&$M_Z$&$M_W$&$m$&$0$&$0$
                        \protect\\
                        \hline
                    \end{tabular}
            \end{minipage}
            \hfill
            \begin{minipage}[c]{.5\textwidth}
                \centering
                \scalebox{0.5}
                {\includegraphics{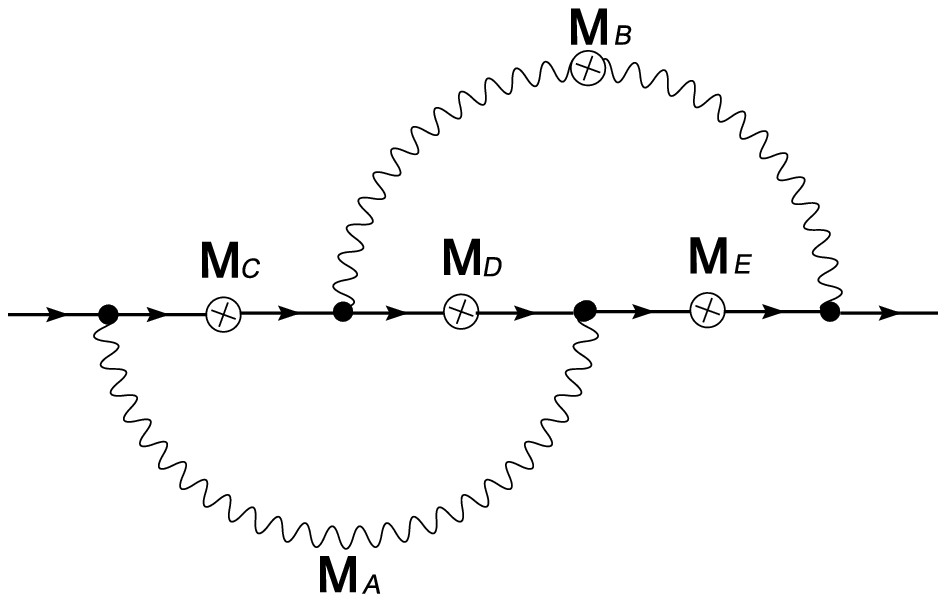}}
            \end{minipage}
    \\
    \\
    \\
        \begin{minipage}[c]{.45\textwidth}
            \begin{tabular}{|c|c|c|c|c|c|}
                \hline
                ${\rm Topology}$&$M_A$&$M_B$&$M_C$&$M_D$&$M_E$
                \protect\\
                \hline
                $2_A$&$M_W$&$M_W$&$M_Z$&$0$&$m$
                \protect\\
                \hline
                $2_B$&$M_W$&$M_W$&$M_H$&$0$&$m$
                \protect\\
                \hline
                $2_C$&$M_Z$&$M_Z$&$M_H$&$0$&$0$
                \protect\\
                \hline
                $2_D$&$M_W$&$M_Z$&$M_W$&$0$&$0$
                \protect\\
                \hline
                $2_E$&$M_W$&$M_W$&$0$&$0$&$m$
                \protect\\
                \hline
                $2_F$&$M_Z$&$M_H$&$M_Z$&$m$&$m$
                \protect\\
                \hline
            \end{tabular}
        \end{minipage}
        \hfill
        \begin{minipage}[c]{.5\textwidth}
            \centering
            \scalebox{0.6}
            {\includegraphics{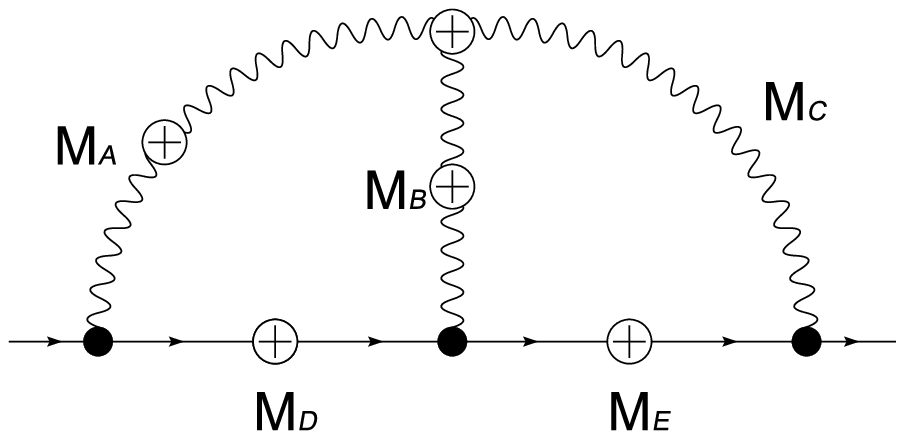}}
        \end{minipage}
    \\
    \\
    \\
        \begin{minipage}[c]{.45\textwidth}
            \begin{tabular}{|c|c|c|c|c|}
                \hline
                ${\rm Topology}$&$M_A$&$M_B$&$M_C$&$M_D$
                \protect\\
                \hline
                $3_A$&$M_W$&$M_W$&$M_W$&$0$
                \protect\\
                \hline
                $3_B$&$0$&$0$&$M_W$&$m$
                \protect\\
                \hline
                $3_C$&$M_Z$&$M_Z/0$&$M_W$&$m$
                \protect\\
                \hline
                $3_D$&$M_W$&$M_W$&$M_Z$&$0$
                \protect\\
                \hline
                $3_E$&$M_W$&$M_W$&$M_H$&$0$
                \protect\\
                \hline
                $3_F$&$M_Z$&$M_Z$&$M_H$&$m$
                \protect\\
                \hline
                $3_G$&$M_Z$&$M_Z$&$M_Z$&$m$
                \protect\\
                \hline
            \end    {tabular}
        \end{minipage}
        \hfill
        \begin{minipage}[c]{.5\textwidth}
            \centering
            \scalebox{0.6}
            {\includegraphics{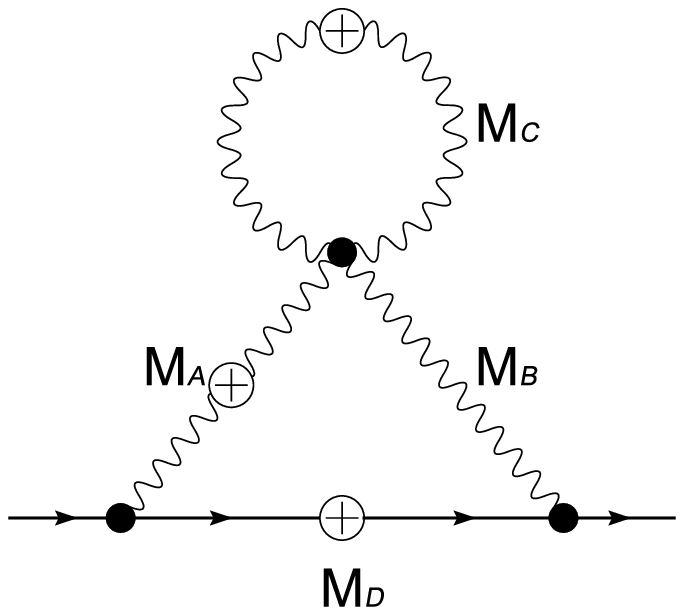}}
        \end{minipage}
    \\
    \\
    \\
        \begin{minipage}[c]{.45\textwidth}
            \begin{tabular}{|c|c|c|c|c|c|}
                \hline
                ${\rm Topology}$&$M_A$&$M_B$&$M_C$&$M_D$&$M_E$
                \protect\\
                \hline
                $4_A$&$M_W$&$M_W$&$M_Z$&$M_W$&$0$
                \protect\\
                \hline
                $4_B$&$M_W$&$M_W$&$M_H$&$M_W$&$0$
                \protect\\
                \hline
                $4_C$&$M_Z$&$M_Z$&$M_H$&$M_Z$&$m$
                \protect\\
                \hline
                $4_D$&$M_W$&$M_W$&$0$&$M_W$&$0$
                \protect\\
                \hline
                $4_E$&$M_Z$&$M_Z$&$M_W$&$M_W$&$m$
                \protect\\
                \hline
                $4_F$&$M_H$&$0$&$M_W$&$M_W$&$m$
                \protect\\
                \hline
                $4_G$&$M_H$&$M_Z$&$M_W$&$M_W$&$m$
                \protect\\
                \hline
                $4_J$&$0$&$0$&$M_W$&$M_W$&$m$
                \protect\\
                \hline
            \end{tabular}
        \end{minipage}
        \hfill
        \begin{minipage}[c]{.5\textwidth}
            \centering
            \scalebox{0.6}
            {\includegraphics{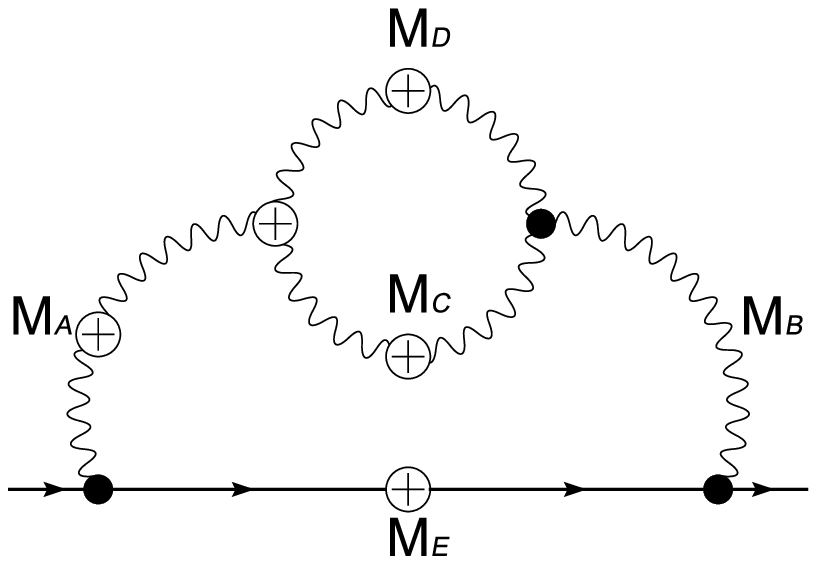}}
        \end{minipage}
    \\
    \\
    \\
        \begin{minipage}[c]{.45\textwidth}
            \begin{tabular}{|c|c|c|c|c|c|}
                \hline
                ${\rm Topology}$&$M_A$&$M_B$&$M_C$&$M_D$&$M_E$
                \protect\\
                \hline
                $5_A$&$M_Z$&$M_Z$&$m$&$m$&$m$
                \protect\\
                \hline
                $5_B$&$M_W$&$M_W$&$0$&$m$&$0$
                \protect\\
                \hline
                $5_C$&$0$&$M_W$&$m$&$0$&$m$
                \protect\\
                \hline
                $5_D$&$M_Z$&$M_W$&$m$&$0$&$m$
                \protect\\
                \hline
                $5_E$&$M_W$&$M_Z$&$0$&$0$&$0$
                \protect\\
                \hline
                $5_F$&$0$&$M_Z$&$m$&$m$&$m$
                \protect\\
                \hline
                $5_G$&$M_Z$&$0$&$m$&$m$&$m$
                \protect\\
                \hline
            \end{tabular}
        \end{minipage}
        \hfill
        \begin{minipage}[c]{.5\textwidth}
            \centering
            \scalebox{0.6}
            {\includegraphics{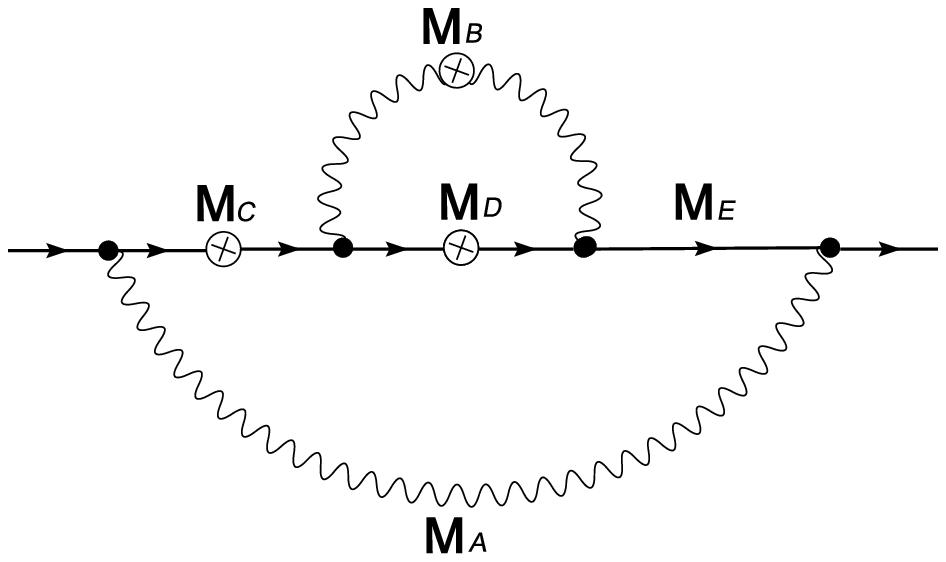}}
        \end{minipage}
\end{tabular}
\caption{Diagram topologies and assignments of masses to their
lines.} \label{TOPOS}
\end{figure}

We divide all two-loop diagrams into five topological types, as
shown in Fig. (\ref{TOPOS}).  Possible insertions of the external
photon vertex are indicated with a circle cross, and wavy lines
stand for either scalar or vector bosons.  Each subset groups
diagrams with similar properties with respect to the asymptotic
operation and thus can be calculated using the same algebraic code
(we use FORM \cite{form3} for most algebraic operations).
 \normalsize
 \linespread{1.0}

Applying the general formula (\ref{GENas}) for the asymptotic
operation  and computing all resulting integrals, we obtain the
following results for the finite parts of various groups of
diagrams:
 \normalsize\linespread{1.0}
\begin{eqnarray}
\label{TsubHiggs}
    \ &&{\rm {T1_{A,B,C}+T2_{A,D,E}+T3_{A,B,C,D,G}+T4_{A,D,E,J}+T5}}
    \nonumber \\
    &=&
    -\frac{\alpha ^2}{384 {\rm c^2} {\rm s^2} \pi^2}  \left\{680 {\rm c^4}-362 {\rm c^2}-363+6
     \left(
        \left[  27-30 {\rm c^2}     \right]
        \ln m^2+    \left[            54 {\rm c^2}-56 {\rm c^4}        \right]
        \ln M_W^2
     \right.
  \right.
\nonumber\\&&
 \left.
    \left.
        +
        \left[
            56 {\rm c^4}-84 {\rm c^2}+27
        \right]
        \ln M_Z^2
     \right)
 \right\}
 -\alpha^2\frac{m^2}{M_W^2}
 \left(
    4.6-0.197\ln \frac{m^2}{M_W^2}
 \right)
\\
\label{T2B} \ {\rm{T2_B}} &=& -\frac{\alpha ^2}{
576\pi^2}\frac{m^2}{M_W^2}\frac{1}{\Delta^4\left(\Delta^2-1\right)}\frac{1}{{\rm
s^4}}
  \left\{
     6\Delta^2
     \left(
        \left[
            2\Delta^6 - 6\Delta^4+3\Delta^2+1
        \right]
        \Phi\left(\frac{\Delta^2}{4}\right)
     \right.
 \right.
 \nonumber\\&&
 \left.
    \left.
        +
        \left[
            4\Delta^6 - 3\Delta^2-4
        \right]
        I^{\rm fin}\left(\Delta^{-2}\right)
     \right)
    +\pi^2
    \left[
        3\Delta^4+\Delta^2-4
    \right]
    + 84\Delta^8+108\Delta^6-111\Delta^4-123\Delta^2
 \right.
\nonumber\\&&
 \left.
    -84+
    \ln \Delta^2
    \left[
        96\Delta^6-78\Delta^2-72
    \right]
    -3\ln^2 \Delta^2
    \left[
        4\Delta^8-8\Delta^6-3\Delta^4+2\Delta^2+8
    \right]
 \right\}
 \\
\label{T2C}\ {\rm{T2_C}} &=& -\frac{\alpha ^2}{
1152\pi^2}\frac{m^2}{M_W^2}\frac{1}{\Lambda^4\left(\Lambda^2-1\right)}\frac{\left(8{\rm
c^4}-12{\rm c^2}+5\right)}{{\rm c^2}{\rm s^4}}
  \left\{
     3\Lambda^2
     \left(
        \left[
            \Lambda^6-3\Lambda^4-6\Lambda^2+8
        \right]
        \Phi\left(\frac{\Lambda^2}{4}\right)
    \right.
  \right.
\nonumber\\&&
  \left.
    \left.
        +
        2\left[
            \Lambda^6- 3\Lambda^2-4
        \right]
        I^{\rm fin}\left(\Lambda^{-2}\right)
     \right)
      +\pi^2
    \left[
        3\Lambda^4+\Lambda^2-4
    \right]
     +3
    \left[
        7\Lambda^8+9\Lambda^6-31\Lambda^4-41\Lambda^2-28
    \right]
 \right.
\nonumber\\&&
 \left.
    +\ln \Lambda^2
    \left[
        24\Lambda^6+18\Lambda^4-78\Lambda^2-72
    \right]
    -3\ln^2 \Lambda^2
    \left[
        \Lambda^8-2\Lambda^6-3\Lambda^4+2\Lambda^2+8
    \right]
 \right\}
\\
\label{T2F}\ {\rm{T2_F}} &=& \frac{\alpha^2}{192\pi^2}\frac{
m^2}{M_W^2}\frac{\Lambda^2}{\left(\Lambda^2-1\right)}\frac{\left({\rm
c^2}-{\rm s^2}\right)}{{\rm {c^2 s^2}}}
 \left\{
     3\left[
        \Lambda^6-7\Lambda^4+14\Lambda^2-8
     \right]
     \Phi\left(\frac{\Lambda^2}{4}\right)
  \right.
\nonumber\\&&
 \left.
     +
     12\Lambda^2\left[
        \Lambda^4-4\Lambda^2+1
     \right]
     I^{\rm fin}\left(\Lambda^{-2}\right)
    +\pi^2\Lambda^2
    \left[
        \Lambda^4-5\Lambda^2+4
    \right]
    +6
    \left[
        7\Lambda^6-21\Lambda^4-20\Lambda^2+6
    \right]
 \right.
\nonumber\\&&
 \left.
    +12\ln \Lambda^2
    \left[
        3\Lambda^4-11\Lambda^2+2
    \right]
    -3\ln^2 \Lambda^2
    \left[
        \Lambda^6-7\Lambda^4+14\Lambda^2-4
    \right]
 \right\}
\\
\label{T3}\ {\rm{T3_{E,F}}} &=& -\frac{\alpha
^2}{2304\pi^2}\frac{m^2}{M_W^2}\Delta^2\frac{1}{{\rm s^4}}
  \left\{
    79-336{\rm c^{2}}+224{\rm c^{4}}
    +72\ln {m^2}
    \left[
        1-3{\rm c^{2}} +2{\rm c^{4}}
    \right]
 \right.
\nonumber\\&&
 \left.
    +6\ln {M_H^2}
    \left[
        7-12{\rm c^{2}}+8{\rm c^{4}}
    \right]
    +30\ln {M_W^2}-12\ln {M_Z^2}
    \left[
        5-12{\rm c^{2}}+8{\rm c^{4}}
    \right]
 \right\}
\\
\label{T4B}\ {\rm{T4_B}} &=&\frac{\alpha ^2}{
2304\pi^2}\frac{m^2}{M_W^2}\frac{1}{\left(\Delta^2-1\right)}\frac{1}{{\rm
s^4}}
  \left\{
     -18
     \left[
        \Delta^8-6\Delta^6+7\Delta^4+4\Delta^2-6
     \right]
     \Phi\left(\frac{\Delta^2}{4}\right)
     - 36\Delta^4
     \left[
        \Delta^4
     \right.
  \right.
\nonumber\\&&
 \left.
    \left.
        -3\Delta^2-2
     \right]
     I^{\rm fin}\left(\Delta^{-2}\right)
    -126\Delta^8+216\Delta^6+635\Delta^4+650\Delta^2-367
    -12\Delta^2\ln \Delta^2
    \left[
        12\Delta^4
    \right.
  \right.
\nonumber\\&&
 \left.
    \left.
        -34\Delta^2-17
    \right]
    +18\ln^2 \Delta^2
    \left[
        \Delta^8-5\Delta^6+4\Delta^4+4\Delta^2
    \right]
    +60\ln {M_W^2}
    \left[
        \Delta^4-4\Delta^2+3
    \right]
 \right\}
\\
\label{T4C}\ {\rm{T4_C}}&=& \frac{\alpha
^2}{576\pi^2}\frac{m^2}{M_W^2}\frac{1}{\left(\Lambda^2-1\right)}\frac{1}{{\rm
s^4}{\rm c^{2}}} \times
  \left\{
    \left(
        \Lambda^2 - 1
    \right)
    \left(
        38\Lambda^2-125+\rm{c^2}(2\rm{c^2}-3)(25\Lambda^2-91)
    \right)
 \right.
\nonumber\\&&
 \left.
    +18\ln {m^2}
    \left(
        1-3{\rm c^2}+2{\rm c^4}
    \right)
    \left(
        \Lambda^4-4\Lambda^2+3
    \right)
    +3\Lambda^2\ln {M_H^2}
    \left(
        \Lambda^2-4
    \right)
    \left(
        1-6{\rm c^2}+4{\rm c^4}
    \right)
 \right.
\nonumber\\&&
 \left.
    -3\ln {M_Z^2}
    \left[
        12-20\Lambda^2+5\Lambda^4+
        \left(
            \rm{4c^4-6c^2}
        \right)
        \left(
            3-8\Lambda^2+2\Lambda^4
        \right)
    \right]
 \right\}
\\
\label{T4F}\ {\rm{T4_F}} &=& -\frac{\alpha^2}{
16\pi^2}\frac{m^2}{M_H^2}\frac{1}{\Delta^2}\frac{1}{{\rm s^2}}
  \left\{
     \left[
        6-7\Delta^2
     \right]
     \Phi\left(\frac{\Delta^2}{4}\right)+
     \Delta^2\left(
        6+\Delta^2
     \right)
     \left(
        \ln \Delta^2-2
     \right)
 \right\}
\\
\label{T4G}
 \ {\rm T4_G}&=& \frac{\alpha^2}{
128\pi^2}\frac{m^2}{M_H^2}\frac{\Lambda^2}{\left(\Lambda^2-1\right)}\frac{\left(4{\rm
c^2}-3\right)}{{\rm s^4}}
  \left\{
     \left[
        14-3\frac{\Delta^2}{\Lambda^2}+2\frac{\left(1-6{\rm
        c^2}\right)}{\Lambda^2}
     \right]
     \Phi\left(\frac{\Delta^2}{4}\right)
 \right.
\nonumber\\&&
 \left.
     +
     \left(
        2\Lambda^2+12{\rm
        c^2}-\Delta^2-18+\frac{4\Delta^2}{\Lambda^2}
     \right)
     \Phi\left(\frac{1}{4{\rm c^2}}\right)
     -\ln \Lambda^2
     \left(
        2\Delta^2+12-\frac{\Delta^4}{\Lambda^2}-2\frac{\Delta^2}{\Lambda^2}
     \right)
 \right \}
 \,,
\end{eqnarray}
In the above expressions we have dropped the divergences (which
cancel in the final sum with the counterterms).  We also express all
masses in units of $\mu=1$ GeV.  The floating point coefficients in
(\ref{TsubHiggs}) were rounded to provide the precision of
$10^{-11}$ of the final result.  They were obtained using
$M_W = 80.423$ GeV and $M_Z = 91.1876$ GeV.  The functions $\Phi(x)$ and $I^{\rm
fin}(x)$ are defined by means of (\ref{Phi1}, \ref{Phi2} and
\ref{M2int}) respectively.  We also use the notation $\Delta \equiv
\frac{M_H}{M_W}$,  $\Lambda \equiv \frac{M_H}{M_Z}$,  ${\rm c}\equiv
\frac{M_W}{M_Z}$, and ${\rm s} \equiv \sqrt{1-{\rm c^2}}$.

The results for topologies $T1_A$, $ T1_B$, $ T1_C$, $ T2_A$, $
T2_D$, $ T2_E$, $ T3_A$, $ T3_B$, $ T3_C$, $ T3_D$, $ T3_G$, $
T4_A$, $ T4_D$, $ T4_E$, $ T4_J$, $ T5$, were collected into a
single formula because the corresponding diagrams do not contain
Higgs propagators and thus can be evaluated without any
assumptions about $M_H$.

\subsection{Renormalization counterterms}
\label{renormalization}

Conceptually, the renormalization procedure for our calculation is
identical to the one we used in Ref. \cite{CKM96}. The only
technical difference is that in our calculation we computed one-loop
massive self-energy diagrams precisely, rather than in an expansion
to the order of $ {\sin^6 \theta _W }$. In this
section we briefly review our renormalization procedure and list
specific expressions for the $W$ and $Z$ self-energies and for other
renormalization constants.

The counter terms for the two-loop $EW$ corrections   are generated
by renormalizing vertices and propagators in one-loop diagrams. In
the non-linear 't Hooft-Feynman gauge there are just three such
diagrams: the Schwinger's QED diagram with a photon loop,  its
analog with the photon replaced by a $Z$ boson, and a diagram with
two $W$ bosons.

The result for Schwinger's diagram in dimensional regularization
is given by
\begin{eqnarray}
\label{Schwinger1} a^{\rm Schw}=\frac{\alpha}{2\pi}
 \left[
    1+\varepsilon
    \left(
        4-\ln m^2
    \right)
 \right]\,.
\end{eqnarray}For the calculation of the counterterms we also need the
contribution of that diagram with one internal muon propagator
squared. It is
\begin{eqnarray}
\label{Schwinger2} a^{\rm Schw}_2=\frac{i \alpha}{2\pi m}
 \left[
    1+\varepsilon
    \left(
        1-\ln m^2
    \right)
 \right]\,.
\end{eqnarray}Using  (\ref{Schwinger1}) and  (\ref{Schwinger2}),
and bearing in mind that since for the calculation of the $EW$
corrections we droped the two-loop photonic corrections to the
Schwinger diagram, and thus have to subtract the photonic
contributions from muon mass and wave function renormalization
constants as well, we find the counterterm to the $a_f$ at the
two-loop level,
\begin{eqnarray}
\label{SchwingerCT} a^{\rm Schw}_{\rm CT}&=&a^{\rm Schw}
 \left[
     \frac{1}{2}
     \left(
        \delta Z^R_{\mu}+\delta Z^L_{\mu}-2\delta Z^{\gamma}_{\mu}
     \right)
     +\Sigma^{'AA}
     \left(
        0
     \right)
 \right]
 -2ia^{\rm Schw}_2
 \left(
    \delta m-\delta m^{\gamma}
 \right)
 \,.
\end{eqnarray}

One-loop contribution of the $Z$-loop diagram  is
\begin{eqnarray}
\label{Zloop1}
 &&a^Z  = \frac{{m^2 }}{{M_Z^2 }}\frac{\alpha }{{4\pi }}
 \left(
    {g_V^2 V + g_A^2 A}
 \right)
 \nonumber\\&&
 V \equiv \frac{1}{3} + \varepsilon
 \left(
    { - \ln m^2  + \frac{2}{3}\ln M_Z^2  - \frac{{11}}{9}}
 \right)
 \nonumber\\&&
 A \equiv  - \frac{5}{3} + \varepsilon
 \left(
    { \ln m^2  + \frac{2}{3}\ln M_Z^2  - \frac{{11}}{9}}
 \right)\,.
\end{eqnarray}
The corresponding diagram with one muon propagator squared gives
\begin{eqnarray}
\label{Zloop2}
 a_2^Z  = \frac{{im^2 }}{{M_Z^2 }}\frac{\alpha }{{4\pi }}
 \left(
    {g_V^2  - g_A^2 }
 \right)
 \left[
    \frac{1}{2} - \varepsilon
    \left(
        {\frac{1}{2}\ln m^2  + 1}
    \right)
 \right]
\end{eqnarray}Combining expressions  (\ref{Zloop1}) and
 (\ref{Zloop2}) we obtain the counterterm generated by the $Z$-loop
diagram,
 \normalsize\linespread{1.0}
\begin{eqnarray}
\label{ZloopCT} a_{CT}^Z  &=&  - 2ia_2^Z \delta m  +
\frac{{m^2}}{{M_Z^2 }}\frac{\alpha }{{4\pi }} \left[
    Vg_V^2
    \left(
        {\frac{{\delta Z_{\mu}^R  + \delta Z_{\mu}^L }}{2} - \frac{{\delta
        M_Z^2 }}{{M_Z^2 }} + 2\frac{{\delta g_V }}{{g_V }}}
    \right)
    +Ag_A^2
    \left(
        \frac{{\delta Z_{\mu}^R  + \delta Z_{\mu}^L }}{2}
    \right.
\right.
\nonumber\\&&
\left.
    \left.
        -\frac{{\delta M_Z^2 }}{{M_Z^2 }} + 2\frac{{\delta g_A }}{{g_A }}
    \right)
    + g_V g_A \frac{{V + A}}{2}
    \left(
        {\delta Z_{\mu}^R  - \delta Z_{\mu}^L }
    \right)
    + \frac{2}{3}\varepsilon
    \left(
        {g_V^2  + g_A^2 }
    \right)
    \frac{{\delta M_Z^2 }}{{M_Z^2 }}
\right]
\end{eqnarray}

Finally, the one-loop contribution of the diagram with two $W$
bosons  is
\begin{eqnarray}
 \label{WW} a^W  = \frac{{m^2 }}{{s^2 M_W^2}}\frac{\alpha }{{4\pi }} \left[
    {\frac{5}{6} + \varepsilon
    \left(
        { - \frac{5}{6}\ln M_W^2  + \frac{{19}}{{36}}}
    \right)}
\right]\,.
\end{eqnarray} Using this expression we obtain a
relatively simple counter term because there is no internal muon
line and there are only left-handed couplings,
\begin{eqnarray}
\label{WWct} a_{CT}^W  = a^W
 \left(
    {2\delta Z_e  - 2\frac{{\delta s }}{{s }} + \delta Z_{\mu}^L  -  \frac{{\delta M_W^2 }}{{M_W^2 }}}
 \right)
 -\frac{5}{6}\frac{{m^2 }}{{s^2 M_W^2 }}\frac{\alpha}{{4\pi }}\varepsilon \frac{{\delta M_W^2 }}{{M_W^2
 }}\,.
\end{eqnarray}
The constants appearing in the expressions
 (\ref{Schwinger1})--(\ref{WWct}) are defined as  follows:
\begin{eqnarray}
\label{gVgA} g_V  &=& \frac{1}{{s c }} \left(
    { -\frac{1}{2} + 2 s^2 }
\right) , \qquad g_A  = \frac{1}{{2 s c }}\,, \nonumber \\
\label{dgVdgAdsw} \delta g_A & =&  - \frac{1}{{2 s c }}
 \left[
    {\delta Z_e  +
    \left(
        {\frac{{s^2}}{{c^2 }} - 1}
    \right)
    \frac{{\delta s }}{{s}}}
 \right],
\nonumber\\
 \delta g_V & =&  - \delta g_A  + 2 \frac{{s}}{{c }}
 \left(
    {\delta Z_e  + \frac{1}{{c^2 }}\frac{{\delta s}}{{s }}}
 \right),
\nonumber\\
 \delta s &=&  - \frac{{c^2}}{{2 s}}
 \left(
    {\frac{{\delta M_Z^2 }}{{M_Z^2 }}- \frac{{\delta M_W^2 }}{{M_W^2 }}}
 \right)
 \,,
\nonumber \\
\label{dZe} \delta Z_e  &=& \frac{{\Sigma ^{'AA}
 \left(
    0
 \right)
 }}{2},\hspace{5mm}\Sigma ^{'AA}
 \left(
    0
 \right)
 = \frac{\alpha }{{4\pi}}
 \left(
    { - \frac{7}{\varepsilon } + 7\ln M_W^2  - \frac{2}{3}}
 \right)\,,
\end{eqnarray}
and $\delta m , \delta Z_{\mu}^{L,R} $ are the muon mass and wave
function renormalization defined as
\begin{eqnarray}
\label{dm} \delta m &=&\delta m^{\gamma}+\delta m^Z+\delta
m^W\,,\hspace{5mm} \delta m^{\gamma}= m\frac{\alpha}{4\pi}
 \left(
    -4-\frac{3}{\varepsilon}+3\ln m^2
 \right)\,,
 \nonumber\\
\delta m^Z
 &=& \frac{\alpha }{{4\pi}}\frac{m}{s^2}
 \left[
    { -\frac{1}{16}+ \frac{1}{8\varepsilon } -\frac{\ln M_Z^2}{8} - \frac{1}{12}\frac{m^2}{M^2_Z}}
 \right]
  \nonumber\\&&
 +\frac{\alpha }{{4\pi}}\frac{m}{c^2}
 \left[
    -\frac{21}{16}- \frac{11}{8\varepsilon } +\frac{11 \ln M_Z^2}{8} + \frac{m^2}{M^2_Z}
    \left(
        -\frac{17}{12}+2\ln\frac{M_Z^2}{m^2}
    \right)
 \right]
 \nonumber\\&&
 +m\frac{\alpha }{{4\pi}}
 \left[
    \frac{5}{2}+ \frac{3}{\varepsilon } -3 \ln M_Z^2+
    \frac{m^2}{M^2_Z}
    \left(
        \frac{8}{3}-4\ln\frac{M_Z^2}{m^2}
    \right)
 \right]\,,
 \nonumber\\
 \delta m^W
 &=& \frac{\alpha }{{4\pi}}\frac{m}{s^2}
 \left[
    { -\frac{1}{8}+ \frac{1}{4\varepsilon } -\frac{\ln M_W^2}{4} + \frac{1}{12}\frac{m^2}{M^2_W}}
 \right]\,,
\end{eqnarray}
and
\begin{eqnarray}
\label{dZ} \delta Z_{\mu}^L &=& \delta Z_{\mu}^L ({\gamma})+\delta
Z_{\mu}^L(Z)+\delta Z_{\mu}^L(W)\,,
 \nonumber\\
 \delta Z_{\mu}^L(\gamma)&=&\delta Z_{\mu}^R(\gamma)=\frac{\alpha}{4\pi}
 \left(
    -4-\frac{3}{\varepsilon}+3\ln m^2
 \right)\,,
 \nonumber\\
 \delta Z_{\mu}^L(Z)&=&\frac{\alpha }{{4\pi}}\frac{1}{16 s^2 c^2}
 \left[
    (2-4s^2)^2
    \left(
        \frac{1}{2}- \frac{1}{\varepsilon } +\ln M_Z^2-\frac{2 m^2}{3 M^2_Z}
    \right)
 \right.
 \nonumber\\ &&\qquad\qquad\qquad
 \left.
    -\frac{2 m^2}{3 M^2_Z}
    \left(
        7-5(1-4 s^2)^2
    \right)
 \right]\,,
\nonumber\\
 \delta Z_{\mu}^L(W)&=&\frac{\alpha}{{4\pi}}\frac{1}{ s^2}
 \left[
    \frac{1}{4}- \frac{1}{2 \varepsilon } +\frac{\ln M_W^2}{2}-\frac{m^2}{3 M^2_W}
 \right]\,,
\nonumber \\
\delta Z_{\mu}^R &=& \delta Z_{\mu}^R ({\gamma})+\delta
Z_{\mu}^R(Z)\,,
 \nonumber\\
 \delta Z_{\mu}^R(Z)&=&\frac{\alpha }{{4\pi}}\frac{1}{16 s^2 c^2}
 \left[
    16s^4
    \left(
        \frac{1}{2}- \frac{1}{\varepsilon } +\ln M_Z^2-\frac{2 m^2}{3 M^2_Z}
    \right)
    -\frac{2 m^2}{3 M^2_Z}
    \left(
        7-5(1-4 s^2)^2
    \right)
 \right]\,,
\end{eqnarray}
and $\delta M^2_W$ and $\delta M_Z^2$ are mass renormalization
constants.

The expressions  (\ref{SchwingerCT}, \ref{ZloopCT}, \ref{WWct})
provide all counter terms necessary to renormalize two-loop
bosonic corrections to $a_{\mu}$. Taking the sum of the finite
parts of these expressions we obtain for the counter terms,
\begin{eqnarray}
\label{CTfull} \ {\rm CT} &=&
 \frac{\alpha ^2}{384 {\rm c^2} {\rm s^2} \pi^2}
  \left\{
    680 {\rm c^4}-362 {\rm c^2}-363
  \right.
\nonumber\\&&
 \left.
  +6
     \left(
        \left[
            27-30 {\rm c^2}
        \right]
        \ln m^2+
        \left[
            54 {\rm c^2}
            -56 {\rm c^4}
        \right]
        \ln M_W^2
        +
        \left[
            56 {\rm c^4}-84 {\rm c^2}+27
        \right]
        \ln M_Z^2
     \right)
 \right\}
\nonumber\\&&
 -\alpha^2\frac{m^2}{M_W^2}
 \left(
    0.74+0.058\ln \frac{m^2}{M_W^2}
    -0.111\Delta^2-0.0134\Delta^2\ln \Delta^2
 \right)
\nonumber\\&&
 -\frac{\alpha ^2}{384 {\rm s^6} \pi^2} \frac{ m^2 }{M_W^2}
 \left\{
    \frac{S^{01}_{\rm fin}(M_W, M_W, M_Z)}{M_W^2}
    \left(
        {\rm -96+475c^2-924c^4+828c^6-256c^8}
    \right)
  \right.
\nonumber\\&&
 \left.
    +
    \left(
        \rm{7-14c^2 + 4c^4}
    \right)
    \left(
         \frac{S^{01}_{\rm fin}(M_W, M_H, M_W)}{M_W^2}
        +
        \left(
            \rm{1+28c^2}
        \right)
        \frac{S^{01}_{\rm fin}(M_Z, M_W, M_W)}{M_W^2}
    \right.
 \right.
\nonumber\\&&
 \left.
    \left.
        +3S^{02}_{\rm fin}(M_W, M_H, M_W)-
        \left(
            \Delta^2-2
        \right)
        S^{12}_{\rm fin}(M_W, M_H, M_W)-S^{22}_{\rm fin}(M_W, M_H, M_W)
   \right.
 \right.
\nonumber\\&&
 \left.
    \left.
            -
            \left(
                \rm{1+48c^2}
            \right)
            S^{22}_{\rm fin}(M_Z, M_W, M_W)
    \right)
     +
    \left(
        \rm{5-16c^2+8c^4}
    \right)
    \times
 \right.
\nonumber\\&&
 \left.
    \left(
        -\rm{c^2} \frac{S^{01}_{\rm fin}(M_Z, M_H, M_Z)}{M_W^2}
        +
            \left(
                \rm{1-4c^2 + 12c^4}
            \right)
            \left(
                S^{22}_{\rm fin}(M_W, M_W, M_Z)
                -S^{12}_{\rm fin}(M_W M_W M_Z)
            \right)
    \right.
 \right.
\nonumber\\&&
 \left.
    \left.
            -3S^{02}_{\rm fin}(M_Z, M_H, M_Z)
            +
            \left(
                \Lambda^2-2
            \right)
            S^{12}_{\rm fin}(M_Z M_H M_Z)+S^{22}_{\rm fin}(M_Z M_H M_Z)
    \right)
 \right.
\nonumber\\&&
 \left.
    +
    \left(
        \rm{96-515c^2+1212c^4-1404c^6+512c^8}
    \right)
    S^{02}_{\rm fin}(M_W, M_W, M_Z)
 \right.
\nonumber\\&&
 \left.
    +\left(
        \rm{-133+82c^2+322c^4-169c^6+21c^8}
    \right)
    S^{02}_{\rm fin}(M_Z, M_W, M_W)
 \right.
\nonumber\\&&
 \left.
    +
    \left(
        \rm{217-346c^2-42c^4+29c^6+7c^8}
    \right)
    S^{12}_{\rm fin}(M_Z, M_W, M_W)
 \right\}\,,
\end{eqnarray}
where, as in expression (\ref{TsubHiggs}), the floating point
coefficients were rounded to provide the precision of $10^{-11}$ of
the result and do not depend on $M_W$ and $M_Z$ chosen within
experimentally accepted interval.  All special functions $S_{a,
b}^{fin}$ are defined in Eq.~(\ref{SEint}).

\subsection{Two-loop corrections to $g-2$}
\label{results}

Our result (not yet in its final form -- see the next section)
for the two-loop bosonic
contribution to $g_\mu-2$ is obtained by adding up
all diagrams, Eqs.~(\ref{TsubHiggs})--(\ref{T4G}),
and  counterterms (\ref{CTfull}),
\begin{eqnarray}
\label{ResFull}
 \ a_{\mu}^{\rm EW\ bos}({\rm
 two-loop})&=&T1_{A,B,C}+T2_{A,D,E}+T3_{A,B,C,D,G}+T4_{A,D,E,J}+T5+T2_{B}
 \nonumber\\&&
 +T2_{C}+T2_{F}+T3_{E,F}+T4_{B}+T4_{C}+T4_{F}+T4_{G}+CT.
\label{eq:result}
\end{eqnarray}

\begin{figure}[!h]
\begin{tabular}{cc}
&
    \centering
    \scalebox{0.75}
    {\includegraphics {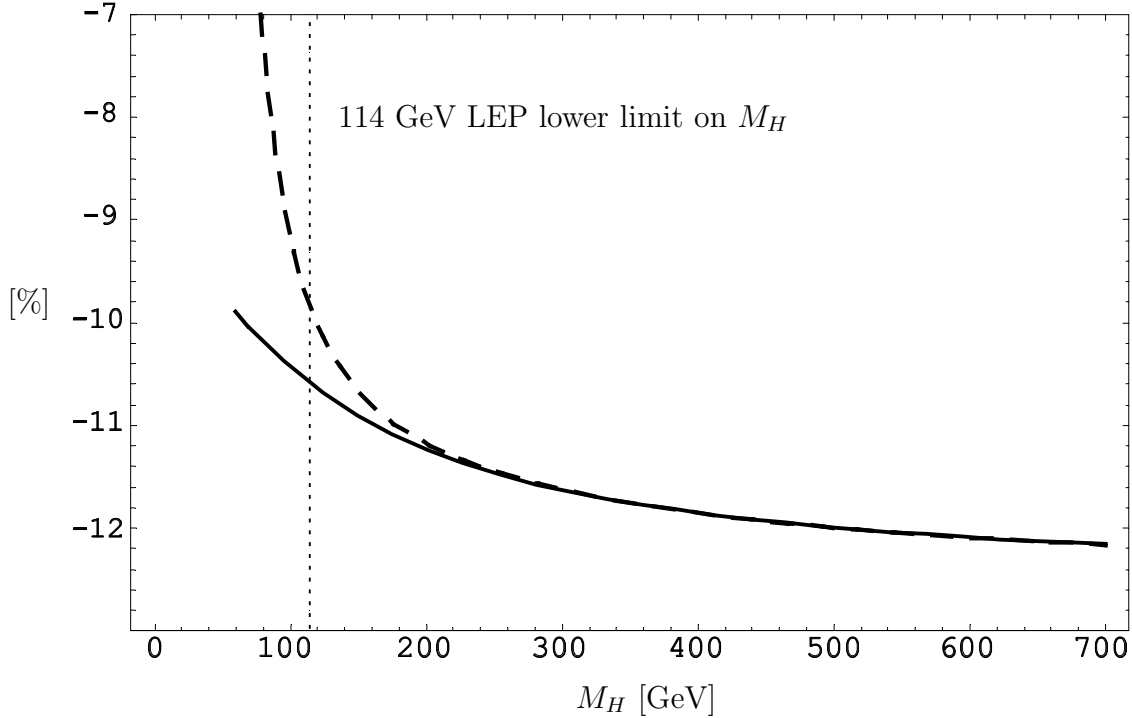}}
\end{tabular}
    \put(-220,-140){\large $M_H$ [GeV]}
    \put(-310,80){\large 114 GeV LEP lower limit on $M_H$}
    \put(-435, 10){\large $[\%]$}
    \caption{$\frac{a_{\mu}^{\rm EW}(\rm two-loop)}{a_{\mu}^{\rm EW}(\rm one-loop)}$
    as a function of $M_H$, expressed in percents. Dashed line represents the result
    from Ref. \cite{CKM96}, solid line is the result of this work.  The vertical dotted line
shows the lower limit for the Higgs boson mass from direct searches, 114 GeV. }
    \label{comparison}
\end{figure}

Its numerical value, normalized to the one-loop correction giving in Eq.~(\ref{oneLoop}),
is plotted in Fig.~(\ref{comparison}) with a solid line,
for a range of the Higgs boson mass  50 GeV
 $\le  M_H  \le $ 700 GeV.  Also plotted, with a dashed line, is the approximate
result of Ref. \cite{CKM96}.
We see that both results coincide well for the Higgs boson heavier than about 200 GeV.
The only difference in the large Higgs mass region is due to an additional approximation
made in Ref. \cite{CKM96}, where the difference of the $W$ and $Z$ masses was treated as a
small perturbation (first four terms in an expansion in $\sin^2\theta_W$ were retained there).
Above $M_H =250 $ GeV, the
relative difference between two results is less than $0.1\%$,
well within the precision of the result in Ref.~\cite{CKM96}.  However, the results differ
strongly in the low Higgs mass region.  The result of Ref.~\cite{CKM96} is not valid in that region
since it was obtained under the assumption $M_{W,Z}\ll M_H$.  We see that while the result
of Ref.~\cite{CKM96} seems to grow strongly for the light $M_H$, the actual bosonic correction (solid
line) remains moderate.

\subsection{Reparametrization in terms of $G_{\mu}$}
\label{reparametrization}

Eq.~(\ref{eq:result}) gives a finite result for the two-loop correction,
expressed in terms   of the
fine structure constant $\alpha$, the $W$ mass, and the weak mixing angle.
Some of the corrections computed
in this way are universal for all weak processes and it is convenient
to include them in the lower order result by expressing it  in
terms of the Fermi constant $G_{\mu}$. This amounts to the substitution, to
be made in all components of Eq.~(\ref{eq:result}),
\begin{eqnarray}
\label{transform}
    \frac{e^2}{8~s^2~M^2_W}~\to~\frac{G_{\mu}}{\sqrt  2}(1-\Delta r)
\end{eqnarray}
with
\begin{eqnarray}
\label{delta}
    \Delta r = 2\delta Z_e-2\frac{\delta s}{s}-\frac{\delta
    M_W^2}{M^2_W}+\frac{\Sigma^W(0)}{M_W^2}+\frac{\alpha}{4\pi~s^2}
    \left(
        6+\frac{7-4s^2}{2s^2}
    \right)
    \ln c^2\,.
\end{eqnarray}
This transformation decreases the central value of the ratio of
the two-loop to one-loop corrections by up to $7\%$ (depending on
the Higgs mass) and, in addition, reduces its uncertainty, since
$G_{\mu}$ has been measured to much better precision than $M_W$
and $s$.

\section{Conclusions}

We have presented a detailed evaluation of the two-loop bosonic corrections
to $g_\mu-2$.  Our result confirms the previous approximate evaluation
in the limit of the heavy Higgs boson \cite{CKM96}.  Our final number for this
correction is
\begin{eqnarray}
 a_{\mu}^{\rm EW \ bos}({\rm two-loop})=(-22.2\pm 1.6)\times 10^{-11}\,,
\label{final}
\end{eqnarray}
where the central value was computed for the Higgs mass of 200
GeV, and the error encompasses the interval 50 GeV
 $\le  M_H  \le $ 700 GeV. This is in agreement with the numerical results of an
exact study in Ref.~\cite{Heinemeyer:2004yq}.  In addition to the
numerical result we have provided here a number of semi-analytic
intermediate results. They give insight into details of our
calculation, allow future checks, and simplify the evaluation of
the correction for any Higgs boson mass.

The uncertainty in Eq.~(\ref{final}) is due to the unknown Higgs mass and is slightly reduced in
comparison with previous estimates based on the approximate evalulation.
The main variation of the correction occurs in the
relatively low Higgs mass range.  Thus, when more stringent limits on $M_H$ are
obtained, that uncertainty will quickly decrease.  The main uncertainty of the
electroweak contributions will then be due to electroweak
hadronic effects, as discussed in
\cite{Czarnecki:2002nt}.

\begin{acknowledgments}
This research was supported by the Natural Sciences and
Engineering Research Council of Canada.  TG gratefully
acknowledges support by the Golden Bell Jar Department of Physics
Graduate Scholarship.
\end{acknowledgments}

%\bibliographystyle{../../../Tables/prsty}
%\bibliography{../../../Tables/phd}

\end{document}